\documentclass[twocolumn]{bmcart}

\RequirePackage{natbib}
\usepackage[utf8]{inputenc} 

\usepackage{algorithm}
\usepackage[noend]{algpseudocode}

\usepackage{amsmath,amsfonts}

\usepackage{tikz}
\usetikzlibrary{bayesnet}

\usetikzlibrary{arrows}

\startlocaldefs
\endlocaldefs

\DeclareUnicodeCharacter{2212}{-}
\begin{document}

\begin{frontmatter}

\begin{fmbox}
\dochead{Research}


\title{Bayesian Gamma-Negative Binomial Modeling of\\ Single-Cell RNA Sequencing Data}


\author[
   addressref={aff1},                   
   email={siamak@tamu.edu}   
]{\inits{S}\fnm{Siamak} \snm{Zamani Dadaneh}}
\author[
   addressref={aff3},
   email={pjdefigueiredo@tamu.edu}
]{\inits{P}\fnm{Paul} \snm{de Figueiredo}}
\author[
   addressref={aff4},
   email={shsze@cse.tamu.edu}
]{\inits{SH}\fnm{Sing-Hoi} \snm{Sze}}
\author[
   addressref={aff2},
   email={mingyuan.zhou@mccombs.utexas.edu}
]{\inits{M}\fnm{Mingyuan} \snm{Zhou}}
\author[
   addressref={aff1},
   corref={aff1},
   email={xqian@ece.tamu.edu}
]{\inits{X}\fnm{Xiaoning} \snm{Qian}}


\address[id=aff1]{
  \orgname{Department of Electrical and Computer Engineering, Texas A\&M University}, 
  \city{College Station},                              
  \cny{USA}                                    
}
\address[id=aff2]{%
  \orgname{McCombs School of Business, The University of Texas at Austin},
  \city{Austin},
  \cny{USA}
}
\address[id=aff3]{%
  \orgname{Department of Microbial Pathogenesis and Immunology, Texas A\&M University },
  \city{College Station},                              
  \cny{USA} 
}
\address[id=aff4]{%
  \orgname{Department of Computer Science and Engineering, Texas A\&M University },
  \city{College Station},                              
  \cny{USA} 
}


\begin{artnotes}
\end{artnotes}



\begin{abstractbox}

\begin{abstract} 
\parttitle{Background}
Single-cell RNA sequencing (scRNA-seq) is a powerful profiling technique at the single-cell resolution. Appropriate analysis of scRNA-seq data can characterize molecular heterogeneity and shed light into the underlying cellular process to better understand development and disease mechanisms. The unique analytic challenge is to appropriately model highly over-dispersed scRNA-seq count data with prevalent dropouts (zero counts), making zero-inflated dimensionality reduction techniques popular for scRNA-seq data analyses. Employing zero-inflated distributions, however, may place extra emphasis on zero counts, leading to potential bias when identifying the latent structure of the data. 
\parttitle{Results}
In this paper, we propose a fully generative hierarchical gamma-negative binomial (hGNB) model of scRNA-seq data, obviating the need for explicitly modeling zero inflation. At the same time, hGNB can naturally account for covariate effects at both the gene and cell levels to identify complex latent representations of scRNA-seq data, without the need for commonly adopted pre-processing steps such as normalization. Efficient Bayesian model inference is derived by exploiting conditional conjugacy via novel data augmentation techniques. 
\parttitle{Conclusion}
Experimental results on both simulated data and several real-world scRNA-seq datasets suggest that hGNB is a powerful tool for cell cluster discovery as well as cell lineage inference.
\end{abstract}




\end{abstractbox}
\end{fmbox}

\end{frontmatter}



\section*{Introduction}
Single-cell RNA sequencing (scRNA-seq) has emerged as a powerful tool for unbiased identification of previously uncharacterized
molecular heterogeneity at the cellular level \cite{shapiro2013single}. This is in contrast to standard bulk RNA-seq techniques \cite{nagalakshmi2008transcriptional}, which measures average gene expression levels within a cell population, and thus ignore tissue heterogeneity. Consideration of cell-level variability of gene expressions is essential for extracting signals from complex heterogeneous tissues \cite{macosko2015highly}, and also for understanding dynamic biological processes, such as embryo development \cite{deng2014single} and cancer \cite{patel2014single}.

A large body of statistical tools developed for scRNA-seq data analysis include a dimensionality reduction step. This leads to more tractable data, from both statistical and computational point of views. Moreover, the noise in the data can be decreased, while retaining the often intrinsically low-dimensional signal of interest. Dimensionality reduction of scRNA-seq data is challenging. In addition to high gene expression variability due to cell heterogeneity, the excessive amount of zeros in scRNA-seq hinders the application of classical dimensionality reduction techniques such as principal component analysis (PCA). For instance, in real-world datasets, it has been reported that the first or second principal components often depend more on the proportion of detected genes per cell (i.e., genes with at least one read) than on the actual biological signal \cite{finak2015mast}.

Several existing computational tools adopt explicit zero-inflation modeling to infer the latent representation of scRNA-seq data. Zero-inflated factor analysis (ZIFA) \cite{pierson2015zifa} extends the framework of probabilistic PCA \cite{tipping1999probabilistic} to the zero-inflated setting, by modeling the excessive zeros using Bernoulli distributed random variables which indicate the dropout event. Zero-inflated negative binomial-based wanted variation extraction (ZINB-WaVE) \cite{risso2018general} directly models the scRNA-seq counts using a zero-inflated negative binomial distribution, while accounting for both gene- and cell-level covariates. It infers the model parameters using a penalized maximum likelihood procedure.

Despite its popularity, using an explicit zero-inflation term may place unnecessary emphasis on the zero counts, leading to complication in discovering the latent representation of scRNA-seq data. In this paper, we propose a hierarchical gamma-negative binomial (hGNB) model to both perform dimensionality reduction and adjust for the effects of the gene- and cell-level confounding factors simultaneously. Exploiting the hierarchical structure, the proposed hGNB model is capable of capturing the high over-dispersion present in the scRNA-seq data. More precisely, we factorize the logit of the negative-binomial (NB) distribution probability parameter to identify latent representation of the data. In addition to factorization, linear regression terms are also included in that logit function to adjust for the impact of covariates.

In hGNB, a gamma distribution with varying rate parameter is used to model the cell dependent dispersion parameter of the NB distribution. The cell-level dispersion serves as a means of representing the prevalence of the dropout events. For instance, cells that are sequenced deeply will naturally include less dropped-out genes with zero counts, and thus this will be reflected in the cell specific dispersion parameter of NB distribution.

We follow a Bayesian framework, similar to bulk RNA-seq setting \cite{dadaneh2017bnp,dadaneh2018bayesian, zamani2018covariate}, and derive closed-form Gibbs sampling update equations for the model parameters of hGNB, by exploiting sophisticated data augmentation techniques. More specifically, we apply the data augmentation technique of \cite{zhou2015negative} (2015)  for the NB distribution, and the Polya-Gamma distributed auxiliary variable technique of \cite{polson2013bayesian} (2013) for the closed-form inference of regression coefficients and also latent factor parameters, removing the need for non-trivial Metropolis-Hastings correction steps \cite{chib1995understanding}. Experimental results on several real-world scRNA-seq datasets demonstrate the superior performance of hGNB to identify cell clusters, especially in complex settings, and also its potential application in cell lineages inference.

\section*{Methods}

\subsection*{hGNB model}

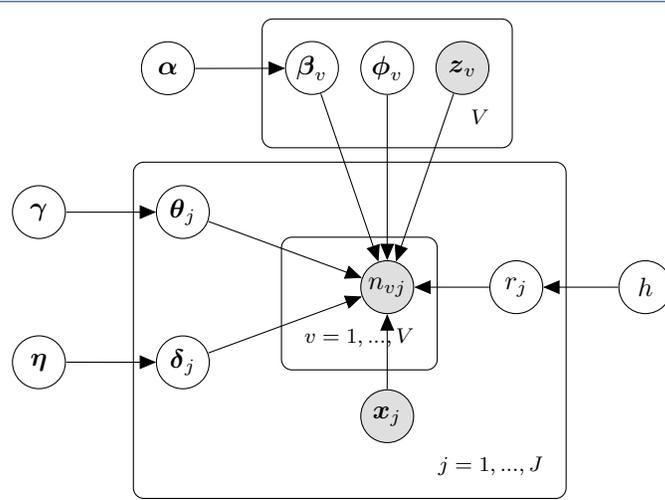
\begin{figure*}[t]
	\centering
	\begin{tikzpicture}[scale=0.95\textwidth/1cm]
	\node[obs] (n) {$n_{vj}$};
	\node[latent,right=of n] (r) {$r_j$};
	\node[latent,above=of n,yshift=1.2cm] (phi) {$\boldsymbol{\phi}_v$};
	\node[obs,above=of n,yshift=1.2cm,xshift=1cm] (z) {$\boldsymbol{z}_v$};
	\node[latent,above=of n,yshift=1.2cm,xshift=-1cm] (beta) {$\boldsymbol{\beta}_v$};
	\node[latent,left=of n,yshift=1cm,xshift=-1cm] (theta) {$\boldsymbol{\theta}_j$};
	\node[latent,left=of n,yshift=-1cm,xshift=-1cm] (delta) {$\boldsymbol{\delta}_j$};
	\node[obs,below=of n] (x) {$\boldsymbol{x}_j$};
	\node[latent,left=of theta,xshift=-0.2cm] (gamma) {$\boldsymbol{\gamma}$}; %
	\node[latent,left=of delta,xshift=-0.2cm] (eta) {$\boldsymbol{\eta}$};
	\node[latent,left=of beta,xshift=-0.2cm] (alpha) {$\boldsymbol{\alpha}$};
	\node[latent,right=of r] (h) {$h$};
	\plate [inner sep=0.3cm] {plate1} {(n)} {$v=1,...,V$};
	\plate [inner sep=0.3cm] {plate2} {(n)(r)(x)(theta)(delta)} {$j=1,...,J$};
	\plate [inner sep=0.3cm] {plate3} {(phi)(beta)(z)} {$V$};
	\edge {phi,theta,beta,x,z,delta,r} {n};
	\edge {gamma} {theta};
	\edge {h} {r};
	\edge {alpha} {beta};
	\edge {eta} {delta};
	\end{tikzpicture}
	\caption{Graphical representation of the hierarchical gamma-negative binomial (hGNB) model.}
	\label{fig:hgnb}
\end{figure*}

In this section we present the hierarchical gamma-negative binomial (hGNB) model for factor analysis of scRNA-seq data. The graphical representation of hGNB is shown in Figure~\ref{fig:hgnb}. The parameters of the hGNB model with their interpretations in the context of scRNA-seq experiments are presented in Table~\ref{tab:model}. Let $n_{vj}$ denote the number of sequencing reads mapped to gene $v \in \{1,...,V\}$ in the cell $j \in \{1,...,J\}$. Under the hGNB model, gene counts are distributed according to a negative binomial (NB) distribution:
\begin{equation}
	n_{vj} \sim \mbox{NB}(r_j,p_{vj}),
\end{equation}
where $r_j$ and $p_{vj}$ are dispersion and probability parameters of NB distribution, respectively. The probability mass function (PMF) of this distribution can be expressed as $f_{N}(n_{vj}) = \frac{\Gamma(n_{vj}+r_j)}{n_{vj}!\Gamma(r_j)} p_{vj}^{n_{vj}} (1-p_{vj})^{r_j}$, where $\Gamma(\cdot)$ is the gamma function.

Data from scRNA-seq experiments exhibit high variability between different cells, even for genes with medium or high levels of expression. To capture this variability, we impose a gamma prior on the cell-level dispersion parameters as
\begin{equation}
	r_j \sim \mbox{Gamma}(e_0,1/h),
	\label{eq:rj}
\end{equation} 
where for simplification, the hyper-parameter $e_0$ is set to 0.01 in our experiments, and the rate $h$ is learned during the Gibbs sampling inference, presented in the following section. This hierarchical prior on the dispersion parameter, enhances the flexibility of NB distribution to capture the high over-dispersion of scRNA-seq counts, without the need for explicit zero-inflation modeling. 

To account for various technical and biological effects common in scRNA-seq technologies, we impose a regression model on the logit of NB probability parameter as
\begin{equation}
	\psi_{vj} = \mbox{logit}(p_{vj}) = \boldsymbol{\beta}_v^T\boldsymbol{x}_j + \boldsymbol{\delta}_j^T\boldsymbol{z}_v + \boldsymbol{\phi}_v^T\boldsymbol{\theta}_j.
	\label{eq:p}
\end{equation}
The three terms in the summation are described below.

In the first term,
$\boldsymbol{x}_j$ is a known vector of $P$ covariates for cell $j$ and $\boldsymbol{\beta}_v$ is the regression-coefficient vector adjusting the effect of covariates on gene $v$. The covariate vector $\boldsymbol{x}_j$ can represent variations of interest, such as cell types, or unwanted variations, such as batch effects or quality control measures. An intercept term can also be included in these cell-level covariates to account for gene dependent baseline expressions.

In the second term, $\boldsymbol{z}_v$ is a vector of $Q$ covariates for gene $v$, representing gene length or GC-content for example \cite{risso2011gc}, and $\boldsymbol{\delta}_j$ is its associated regression-coefficient vector. We also include a fixed intercept element in $\boldsymbol{z}_v$ to account for cell-specific expressions, such as the size factors representing differences in sequencing depth.

In the third term, $\boldsymbol{\phi}_v^T\boldsymbol{\theta}_j$ corresponds to the latent factor representation of the count $n_{vj}$, after accounting for the effects of gene- and cell-level covariates. More precisely, the unknown $K \times 1$ vector $\boldsymbol{\phi}_v$ contains the factor loading parameters which determine the association between genes and latent factors. Moreover, the unknown $K \times 1$ vector $\boldsymbol{\theta}_j$ encodes the popularity of the $K$ factors in the expression of cell $j$.

We place independent zero-mean normal distributions on the components of the regression coefficient parameters $\boldsymbol{\beta}_v$ and $\boldsymbol{\delta}_j$ as
\begin{eqnarray}
	\boldsymbol{\beta}_v &\sim& \prod_{p=1}^{P} \mbox{N}(\beta_{vp};0,\alpha_p^{-1}),\nonumber\\
	\boldsymbol{\delta}_j &\sim& \prod_{q=1}^{Q} \mbox{N}(\delta_{jq};0,\eta_q^{-1}),
	\label{eq:reg}
\end{eqnarray}
where $\alpha_p$ and $\eta_q$ are precision parameters of the normal distributions and gamma priors are imposed on them. These priors are known as automatic relevance determination (ARD), which are effective tools for pruning large numbers of irrelevant covariates \cite{wipf2008new,tipping2001sparse}. In addition, by assuming identical precision for components of the regression coefficients
across all genes or samples, hGNB borrows statistical strengths to infer these precision parameters.

We impose independent normal priors on latent factor loading and score parameters $\boldsymbol{\phi}_v$ and $\boldsymbol{\theta}_j$:
\begin{eqnarray}
\boldsymbol{\phi}_v &\sim& \mbox{N}(\boldsymbol{\phi}_v;0,I_K),\nonumber\\
\boldsymbol{\theta}_j &\sim& \prod_{k=1}^{K} \mbox{N}(\theta_{jk};0,\gamma_k^{-1}).
\label{eq:fac}
\end{eqnarray}
Note that the posterior for these terms is not generally independent or normal, but accounts for the statistical
dependence as reflected in the data.

We complete the model by imposing a gamma prior on the precision parameters of normal distributions, and also the rate parameter of gamma distributions. Specifically, throughout the experiments, we set both the shape and rate of these gamma priors to 0.01. 

\begin{table*}[h]
	\centering
	\caption{Parameters of the hierarchical gamma-negative binomial (hGNB) model and their interpretations in the context of scRNA-seq data. The inputs of hGNB are gene counts $n_{vj}$ and vector of cell- and gene-level covariates $\mathbf{x}_j$ and $\mathbf{z}_v$.}
	\small
	\begin{tabular}{ c| c | c}
		Parameter & Constraint & Interpretation \\
		\hline \hline
		$r_j$ & $r_j > 0$ & expression heterogeneity of genes in sample $j$\\
		\hline
		$\phi_{vk}$ & $\sum_{v=1}^{V} \phi_{vk}=1$, $\phi_{vk}>0$ & gene-latent factor association \\
		\hline
		$\theta_{jk}$ & $\theta_{kj} > 0$ & popularity of factor $k$ in sample $j$ \\
		\hline
		$\beta_{vp}$ & $\beta_{vp} \in \mathbb{R}$ & impact of cell covariate $p$ on expression
		of gene $v$\\
		\hline
		$\delta_{jq}$ & $\beta_{vp} \in \mathbb{R}$ & impact of gene covariate $q$ on expression
		of cell $j$
		
	\end{tabular}
	\label{tab:model}
\end{table*}

\subsection*{Inference via Gibbs sampling}
In this section, we provide an efficient inference algorithm that adopts data augmentation techniques tailored to our hGNB model. Algorithm~1 summarizes all the steps in the Gibbs sampling algorithm.

\paragraph{Sample dispersion parameter} We start with the data augmentation technique developed for inferring the NB dispersion parameter \cite{zhou2015negative}. More precisely, the negative binomial random variable $n \sim \text{NB}(r,p)$ can be generated from a compound Poisson distribution as 
\begin{equation}
n = \sum_{t=1}^{\ell} u_t, \;\; u_t \sim \text{Log}(p), \;\; \ell \sim \text{Pois}(-r\ln (1-p)), \nonumber
\end{equation}
where $u \sim \text{Log}(p)$ corresponds to the logarithmic random variable \cite{johnson2005univariate}, with the PMF $f_U(u) = -\frac{p^u}{u\ln(1-p)}$, $u \in \{1,2,...\}$. As shown in \cite{zhou2015negative}, given $n$ and $r$, the distribution of $\ell$ is a Chinese Restaurant Table (CRT) distribution, $(\ell | n,r) \sim \text{CRT}(n,r)$, which can be generated  as $\ell = \sum_{t=1}^{n} b_t, b_t\sim \text{Bernoulli}(\frac{r}{r+t-1})$.

Utilizing this augmentation technique, for each observed count $n_{vj}$, an auxiliary count is sampled as
\begin{equation} \label{eq:crt}
(\ell_{vj}|-) \sim \mbox{CRT}(n_{vj},r_j).
\end{equation}
Using gamma-Poisson conjugacy, the cell-dependent dispersion parameters are updated as
\begin{equation} \label{eq:rji}
(r_j | -) \sim \mbox{Gamma} \Big( e_0+ \sum_v \ell_{vj}, \frac{1}{h - \sum_v \ln(1-p_{vj})} \Big).
\end{equation}

\paragraph{Sample regression coefficients} For the regression coefficients modeling potential covariate effects, the lack of conditional conjugacy precludes immediate closed-form inference. Therefore we adopt another data augmentation technique, specifically designed for hGNB, to infer the regression coefficients $\boldsymbol{\beta}_v$ and $\boldsymbol{\delta}_j$, relying on the Polya-Gamma (PG) data augmentation~\cite{LGNB_ICML2012,polson2013bayesian}. 

Denote $\omega_{vj}$ as a random variable drawn from the PG distribution as $\omega_{vj} \sim \text{PG}(n_{vj}+r_j,0).$ 
Since  $\mathbb{E}_{\omega_{vj}}[\exp(- \omega_{vj} \psi_{vj}^2/2)]=\cosh^{(n_{vj}+r_j)}(\psi_{vj}^2/2)$,  
the likelihood of $\psi_{vj}$ in~(\ref{eq:p}) can be expressed as
\begin{align}
\mathcal{L}(\psi_{vj}) &\propto \frac{(e^{\psi_{vj}})^{n_{vj}}}{(1+e^{\psi_{vj}})^{n_{vj}+r_j}} \nonumber\\
& \propto \exp\Big(\frac{n_{vj}-r_j}{2} \psi_{vj}\Big) \mathbb{E}_{\omega_{vj}}[\exp(- \omega_{vj} \psi_{vj}^2/2)].
\label{eq:lik}
\end{align}
Exploiting the exponential tilting of the PG distribution in \cite{polson2013bayesian}, we draw 
$\omega_{vj}$ as 
\begin{equation}\label{eq:omega}
(\omega_{vj}|-) \sim \text{PG}(n_{vj}+r_j,\psi_{vj}).
\end{equation}
Given the values of the auxiliary variables $\omega_{vj}$ for $j=1,...,J$ and the prior in~(\ref{eq:reg}), the conditional posterior of $\boldsymbol{\beta}_v$ can be updated as
\begin{equation}
(\boldsymbol{\beta}_v|-) \sim \mbox{N}(\mu_v^{(\beta)},\Sigma_v^{(\beta)}),
\label{eq:beta}
\end{equation}
where $\Sigma_v^{(\beta)} = \Big( \mbox{diag}(\alpha_1,...,\alpha_P)+\sum_{j} \omega_{vj} \boldsymbol{x}_j \boldsymbol{x}_j^T \Big)^{-1}$ and $\mu_v^{(\beta)} = \Sigma_v^{(\beta)} \Big[ \sum_{j} \big(\frac{n_{vj}-r_j}{2} - \omega_{vj} (\boldsymbol{\delta}_j^T\boldsymbol{z}_v + \boldsymbol{\phi}_v^T\boldsymbol{\theta}_j) \big) \boldsymbol{x}_j \Big]$. \\

A similar procedure can be followed to derive the conditional updates for cell-level regression coefficients as
\begin{equation}
(\boldsymbol{\delta}_j|-) \sim \mbox{N}(\mu_j^{(\delta)},\Sigma_j^{(\delta)}),
\label{eq:delta}
\end{equation}
where $\Sigma_j^{(\delta)} = \Big( \mbox{diag}(\eta_1,...,\eta_Q)+\sum_{v} \omega_{vj} \boldsymbol{z}_v \boldsymbol{z}_v^T \Big)^{-1}$ and $\mu_j^{(\delta)} = \Sigma_j^{(\delta)} \Big[ \sum_{v} \big(\frac{n_{vj}-r_j}{2} - \omega_{vj} (\boldsymbol{\beta}_v^T\boldsymbol{x}_j + \boldsymbol{\phi}_v^T\boldsymbol{\theta}_j) \big) \boldsymbol{z}_v \Big]$. 

\paragraph{Sample latent factor parameters} Using the likelihood function in (\ref{eq:lik}) and the priors in (\ref{eq:fac}), we can derive closed-form update steps for factor loading and score parameters. More specifically, the full conditional for factor loading $\boldsymbol{\phi}_v$ is a normal distribution:
\begin{equation}
(\boldsymbol{\phi}_v|-) \sim \mbox{N}(\mu_v^{(\phi)},\Sigma_v^{(\phi)}),
\label{eq:phi}
\end{equation}
where $\Sigma_v^{(\phi)} = \Big( I_K + \sum_{j} \omega_{vj} \boldsymbol{\theta}_j \boldsymbol{\theta}_j^T \Big)^{-1}$ and $\mu_v^{(\phi)} = \Sigma_v^{(\phi)} \Big[ \sum_{j} \big(\frac{n_{vj}-r_j}{2} - \omega_{vj} (\boldsymbol{\beta}_v^T\boldsymbol{x}_j + \boldsymbol{\delta}_j^T\boldsymbol{z}_v) \big) \boldsymbol{\theta}_j \Big]$. 

The full conditional for factor score $\boldsymbol{\theta}_j$ is also a normal distribution:
\begin{equation}
(\boldsymbol{\theta}_j|-) \sim \mbox{N}(\mu_j^{(\theta)},\Sigma_j^{(\theta)}),
\label{eq:theta}
\end{equation}
where $\Sigma_j^{(\theta)} = \Big( \mbox{diag}(\gamma_1,...,\gamma_K) + \sum_{v} \omega_{vj} \boldsymbol{\phi}_v \boldsymbol{\phi}_v^T \Big)^{-1}$ and $\mu_j^{(\theta)} = \Sigma_j^{(\theta)} \Big[ \sum_{v} \big(\frac{n_{vj}-r_j}{2} - \omega_{vj} (\boldsymbol{\beta}_v^T\boldsymbol{x}_j + \boldsymbol{\delta}_j^T\boldsymbol{z}_v) \big) \boldsymbol{\phi}_v \Big]$. 

\paragraph{Sample precision and rate} The precision parameters of normal distributions in (\ref{eq:reg}) and (\ref{eq:fac}) can be updated using the normal-gamma conjugacy:
\begin{eqnarray}
	\alpha_p &\sim& \mbox{Gamma} \big( e_0+V/2, \frac{1}{f_0 + \sum_{v=1}^{V} \beta_{vp}/2} \big), \nonumber\\
	\eta_q &\sim& \mbox{Gamma} \big( e_0+J/2, \frac{1}{f_0 + \sum_{v=1}^{V} \delta_{jq}/2} \big), \nonumber\\
	\gamma_k &\sim& \mbox{Gamma} \big( e_0+J/2, \frac{1}{f_0 + \sum_{v=1}^{V} \theta_{jk}/2} \big).
	\label{eq:hyp}
\end{eqnarray}

Finally, the rate of gamma distribution in (\ref{eq:rj}) can be updated using the gamma-gamma conjugacy with respect to the rate parameter:
\begin{equation}
	h \sim \mbox{Gamma} \big( e_0(1+J), \frac{1}{f_0 + \sum_{j=1}^{J} r_j} \big).
\end{equation}

\begin{algorithm}[t]
	\caption{hGNB model inference}\label{alg:gibbs}
	\textbf{Inputs}: scRNA-seq counts, design matrix of covariate effects, $N$\\
	\textbf{Outputs}: gene module membership matrix
	\begin{algorithmic}
		\State \textit{Initialize} model parameters
		\State \# Do Gibbs sampling:
		\For {$iter=1$ to $N$}
		\State Sample $\ell_{vj}$ using the CRT distribution (eq. (\ref{eq:crt}))
		\State Update $r_{j}$ using the gamma-Poisson conjugacy (eq. (\ref{eq:rji}))
		\State Sample auxiliary variables $\omega_{vj}$, using the PG distribution (eq. (\ref{eq:omega}))
		\State Update cell- and gene-level regression coefficients (eq. (\ref{eq:delta}),(\ref{eq:beta}))
		\State Update factor loadings and scores (eq. (\ref{eq:phi}),(\ref{eq:theta}))
		\State Update $\alpha_p$, $\eta_q$ and $\gamma_k$ (eq. (\ref{eq:hyp})) 
		\EndFor \textbf{end for}
		\State 
	\end{algorithmic}
\end{algorithm}

\section*{Results}
We evaluate our hGNB model on four different sets of real-world scRNA-seq data from different platforms, and compare its performance to those of principal component analysis (PCA), ZIFA \cite{pierson2015zifa}, and ZINB-WaVE \cite{risso2018general}. In the following, We briefly describe these scRNA-seq datasets. To pre-process these datasets when needed, we followed the same procedures as in \cite{risso2018general}.

\textbf{V1 dataset}. This dataset characterizes more than 1600 cells from the primary visual cortex (V1) in adult male mice, using a set of established Cre lines \cite{tasic2016adult}. A subset of three Cre lines, including Ntsr1-Cre, Rbp4-Cre, and Scnn1a-Tg3-Cre, that respectively label layer 4, layer 5, and layer 6 excitatory neurons were selected. We only retained 285 cells that passed the authors' quality control (QC) filters. The dimensionality reduction methods were only applied to the 1000 most variable genes.

\textbf{S1/CA1 dataset}. This dataset characterizes 3005 cells from the primary somatosensory cortex (S1) and the hippocampal CA1 region, using the Fluidigm C1 microfluidics cell capture platform followed by Illumina sequencing \cite{zeisel2015cell}. Gene expression is quantified by UMI counts.

\textbf{mESC dataset}. This dataset includes the transcriptome measurement of 704 mouse embryonic stem cells (mESCs), across three culture conditions (serum, 2i, and a2i), using the Fluidigm C1 microfluidics cell capture platform followed by Illumina sequencing \cite{kolodziejczyk2015single}. We excluded the samples that did not pass the authors's QC filters, resulting in a total of 169 serum cells, 141 2i cells, and 159 a2i cells. The dimensionality reduction methods were only applied to the 1000 most variable genes.

\textbf{OE dataset}. This data characterizes 849 FACS-purified cells from the mouse OE, using the Fluidigm C1 microfluidics cell capture platform followed by Illumina sequencing \cite{fletcher2017deconstructing}. We followed the filtering procedure of \cite{perraudeau2017bioconductor}, and filtered the cells that exhibited poor sample quality, retaining a total of 747 cells.

For all datasets, hGNB was run using 2000 MCMC iterations, where after the first 1000 burn-in iterations, the posterior samples with the highest likelihood were collected as the point estimates of model parameters corresponding to latent factors. In the dimensionality reduction analysis below, following \cite{risso2018general}, for S1/CA1 dataset we set the number of latent factors $K=3$, and for V1 and mESC we set $K=2$.

\subsection*{Goodness-of-fit of hGNB model}
\begin{figure}[t]
	\centering
	\includegraphics[width=0.45\textwidth]{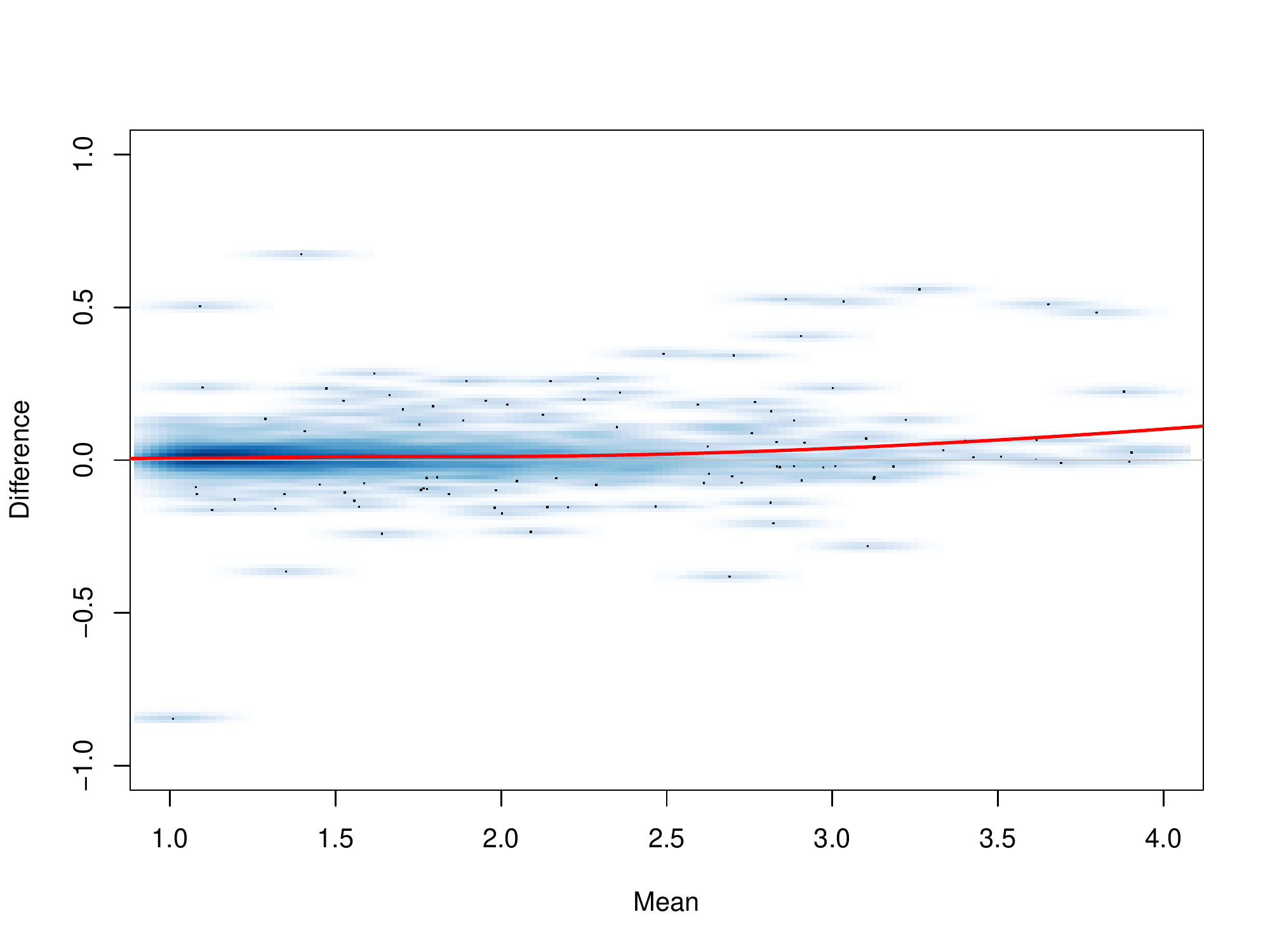}
	\caption{Mean-difference (MD) plot for S1/CA1 dataset. The solid red line represents the local regression fit to the data}
	\label{fig:md}
\end{figure}

We have examined the goodness-of-fit of hGNB model on V1, S1/CA1 and mESC datasets, using the mean-difference (MD) plots. Figure~\ref{fig:md} shows the MD plot for the S1/CA1 dataset, where the y-axis is the difference between observed counts and the expected counts under hGNB, and x-axis is the average of these two sets of counts. The solid red line in this figure, which represents the local regression fit \cite{shyu2017local} to the data, resides near zero for various average levels. This supports the good fit of hGNB model to the highly over-dispersed scRNA-seq data. Similar trends are observed for V1 and mESC datasets (supplementary materials).

\subsection*{Capturing zero-inflation}
Next we evaluate the performance of hGNB on simulated data based on the zero-inflated NB distribution of \cite{risso2018general} to show that hGNB faithfully captures zero inflation without the need of explicit zero-inflation modeling. Specifically, the capability of hGNB to recover true clustering structure of cells under three zero-count prevalence levels with two different total numbers of cells. The parameters of the simulating zero-inflated model were learned based on the S1/CA1 dataset. Genes that did not have at least five reads in at least five cells were filtered out and 1000 genes were then sampled at random for each dataset. The number of latent factors was set to $K=2$. To simulate cell clustering, a $K$-variate Gaussian mixture distribution with three components was fitted to the inferred factor score parameters, and then for each simulated dataset, factor scores were generated from $K$-variate Gaussian distributions. By adjusting the value of regression coefficients in the zero-inflation term of ZINB-WaVE model, we generated synthetic datasets with three levels of zero-count percentages as 40\%, 60\% and 80\% (for details refer to \cite{risso2018general}). The number of cells were set to $J=100$ and $J=1000$. For each scenario, including cell numbers and zero-count prevalence (sparsity) levels, we simulated 10 datasets.

We evaluate the performance of our method for the clustering task based on the average silhouette width measure. The silhouette width $s_j$ of sample $j$ is defined as 
$$ s_j = \frac{b_j-a_j}{\max \{a_j,b_j\}}, $$ 
where $a_j$ is the average distance between sample $j$ and all samples in the cluster that it belongs to, and $b_j$ is the minimum average distance between sample $j$ and samples in other clusters.

\begin{figure}[t]
	\centering
	\includegraphics[width=0.45\textwidth]{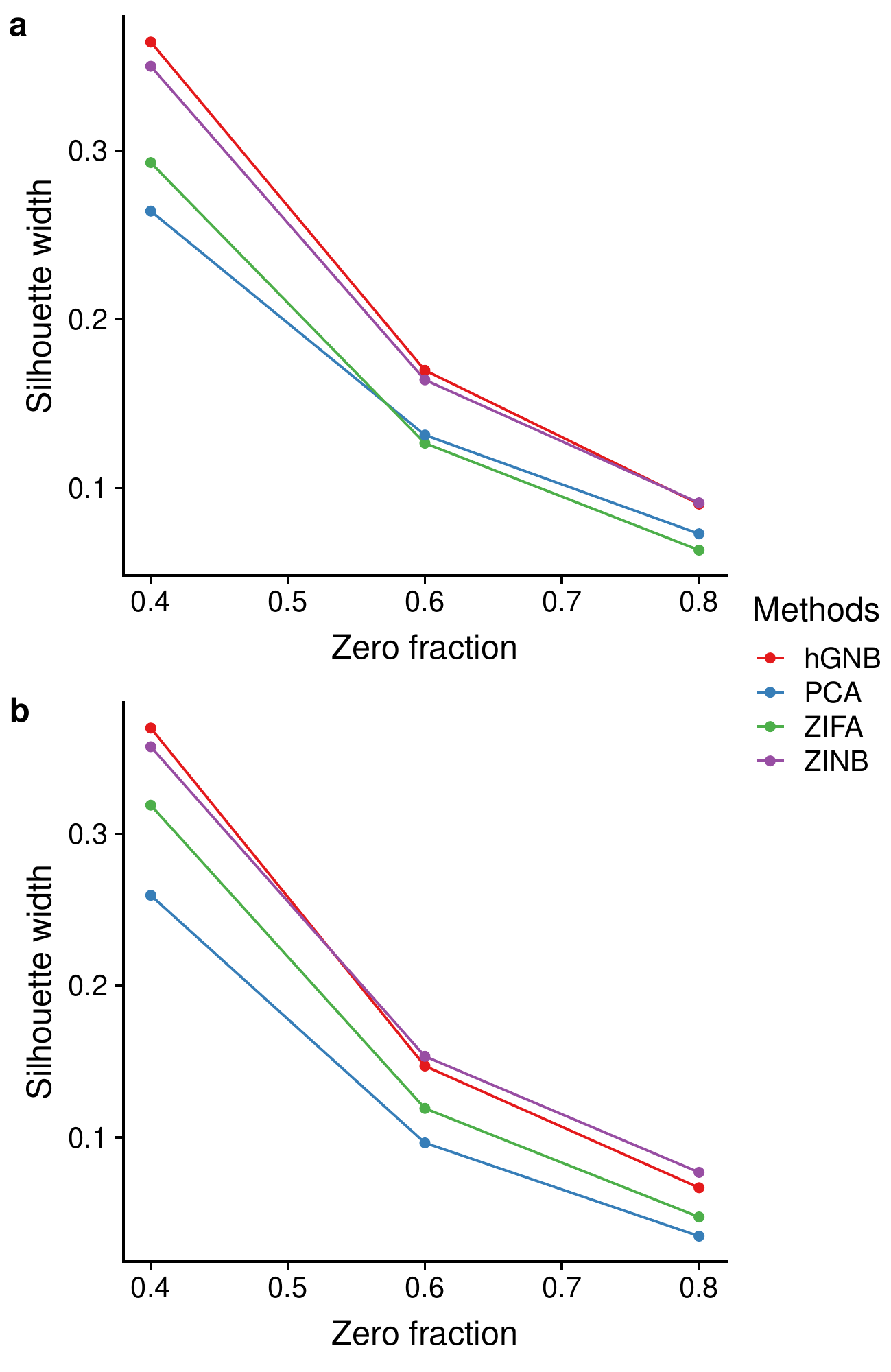}
	\caption{(\textbf{a})$J=100$, (\textbf{b})$J=1000$. Performance of different methods based on recovering the true cell clusters in synthetic data based on S1/CA1 dataset. Zero-inflated NB model of ZINB-WaVE is used to simulate scRNA-seq data.}
	\label{fig:sim}
\end{figure}

Figure~\ref{fig:sim} shows the clustering average silhouette width based on the above simulation setup, for different zero-count prevalence levels and cell numbers. In the setting with small sample size, for 40\% and 60\% zero fractions, hGNB has the best clustering silhouette width, and for the 80\% zero fraction its performance is identical to that of ZINB-WaVE. In the setting with moderate sample size, hGNB has the best clustering silhouette width for 40\% zero fraction, and for 60\% and 80\% zero fractions it closely follows the performance of ZINB-WaVE. This suggests that the hierarchical structure of hGNB equips it with the capacity to capture highly over-dispersed count data, even though an explicit zero-inflation term is not included in its model. Also, ZINB-WaVE requires large enough samples to have robust inference results due to the introduction of zero-inflation terms in its model. Finally, ZIFA and PCA have the worst performance, as they normalize the data before learning its latent representation.

\subsection*{Dimensionality reduction}
\begin{figure*}[t]
	\centering
	\includegraphics[width=0.95\textwidth]{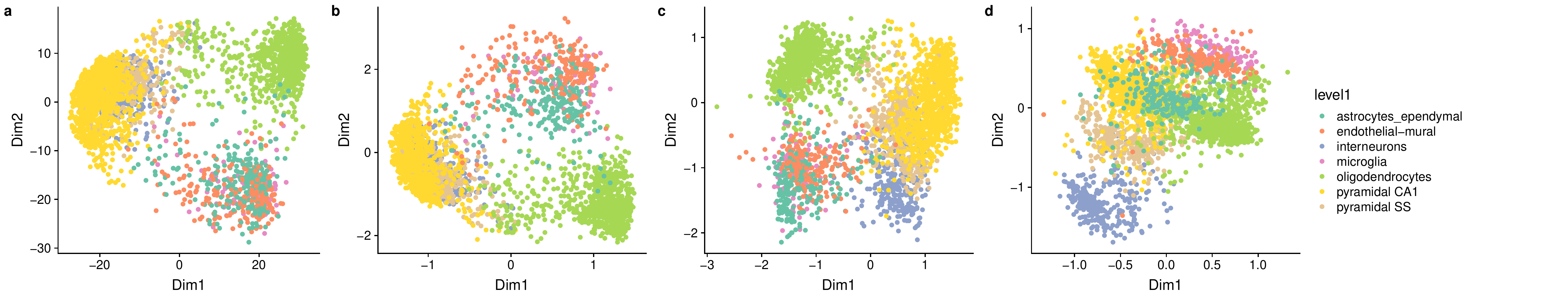}
	\caption{Low-dimensional representations of the S1/CA1 dataset. Panels correspond to (\textbf{a}) PCA (on total-count normalized data), (\textbf{b}) ZIFA (on total-count normalized data), (\textbf{c}) ZINB-WaVE, and (\textbf{d}) hGNB.}
	\label{fig:vis}
\end{figure*}

We applied hGNB to the three scRNA-seq datasets, V1, S1/CA1 and mESC, to assess its power to separate cell clusters in the low dimensional space, and compared it to PCA, ZIFA, and ZINB-WaVE methods. Figure~\ref{fig:vis} illustrates the projected scRNA-seq expression of profiled cells in the two-dimensional space for S1/CA1 dataset. The proposed hGNB model provides more biologically meaningful latent representations of scRNA-seq gene expressions for S1/CA1 cells, especially compared to PCA and ZIFA that do not model the counts directly. Furthermore, hGNB leads to more separated clusters of cells in the two-dimensional space, compared to ZINB-WaVE. Specifically, hGNB distinguishes {\tt microglia} from {\tt endothelial−mural} cells, while ZINB-WaVE fails to accomplish this task.

To examine the dimensionality reduction results more carefully, we used the average silhouette width as a measure of goodness for clustering.

\begin{figure*}[h]
	\centering
	\includegraphics[width=0.95\textwidth]{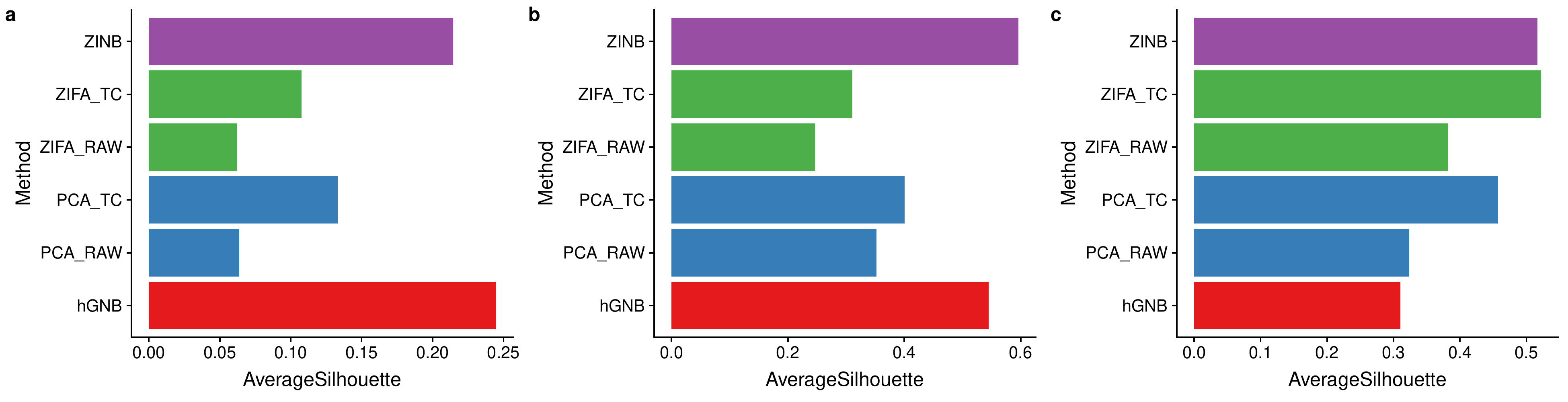}
	\caption{Average silhouette width in scRNA-seq datasets (\textbf{a}) S1/CA1, (\textbf{b}) mESC, and (\textbf{c}) V1. Silhouette widths were
		computed in the low-dimensional space, using the groupings provided by the authors of the original publications. PCA and ZIFA were applied with both unnormalized (RAW) data and after total count (TC) normalization.}
	\label{fig:sil}
\end{figure*}

Figure~\ref{fig:sil} shows the average silhouette width of different methods on V1, S1/CA1, and mESC datasets. For PCA and ZIFA, the results on both raw counts and normalized counts are included in this figure. For S1/CA1 dataset, which has the highest number of clusters, the proposed hGNB method outperforms all other methods in terms of clustering average silhouette. For mESC dataset, performance of hGNB is comparable to ZINB-WaVE, and it is significantly better than PCA and ZIFA. For V1 dataset, however, we observe that hGNB, besides PCA applied to raw counts, possess the lowest average silhouette. By further examination of the latent representations of cells for this dataset (supplementary materials), we observe that all methods split the {\tt Rbp4−Cre\_KL100} cells into two clusters, one of them located near {\tt Scnn1a−Tg3−Cre} cells, suggesting the presence of batch effects, which have led to confounding of latent representations \cite{risso2018general}.

\subsection*{Identification of developmental lineages}
In addition to characterization of cell types, we further demonstrate the capability of hGNB to derive novel biological insights, by analyzing a set of cells from the mouse olfactory epithelium (OE). The samples were collected to identify the developmental trajectories that generate olfactory neurons (mOSN), sustentacular cells (mSUS), and microvillous cells (MV) \cite{fletcher2017deconstructing}.

We first performed dimensionality reduction on the OE dataset by applying hGNB with $K=50$. Next, we clustered the cells using the low-dimensional factor score parameters $\theta_{kj}$. More specifically, the resampling-based sequential ensemble clustering (RSEC) framework implemented in the {\tt RSEC} function from the Bioconductor R package {\tt clusterExperiment} \cite{purdom2017clusterexperiment} was applied to factor scores, leading to identification of 14 cell clusters. The correspondence between the detected clusters and the underlying biological cell types is presented in Table~\ref{tab:oe}. In addition to these already known cell clusters in OE, hGNB is able to detect new clusters, potentially offering novel biological insights.

\begin{table}[t]
	\centering
	\caption{Correspondence between identified clusters and cell types in OE dataset.}
	\small
	\begin{tabular}{| c| c |  }
		\hline
		Cell Type & Clusters \\
		\hline \hline
		GBC & cl4,cl9 \\
		mSUS & cl2,cl3,cl5,cl11 \\
		mOSN & cl8,cl12,cl3 \\
		Immature Neurons & cl10 \\
		MV & cl14\\
		\hline
	\end{tabular}
	\label{tab:oe}
\end{table}

\begin{figure}[!ht]
	\centering
	\includegraphics[width=0.45\textwidth]{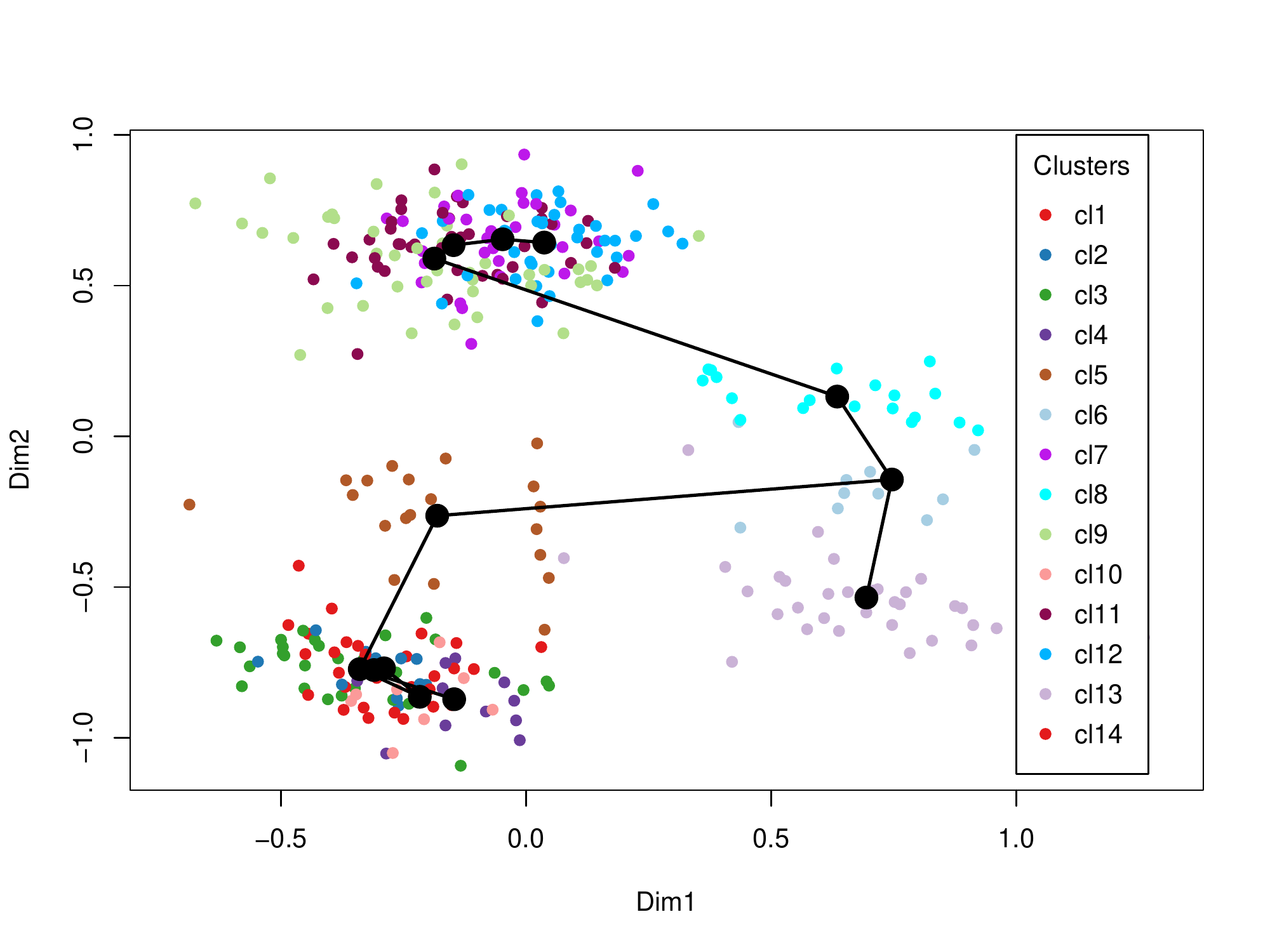}
	\caption{Lineage inference on the OE dataset. The low dimensional data representation derived by hGNB were used to cluster cells by {\tt RSEC}. The minimum spanning tree (MST) of the derived clusters constructed by {\tt slingshot} is also displayed.} 
	\label{fig:lin}
\end{figure}

We further investigated the potential benefit of using the learned latent representation by our proposed hGNB model to infer branching cell lineages and order cells by developmental progression along each lineage. To infer the global lineage structure (i.e., the number of lineages and where they branch), a minimum spanning tree (MST) was constructed on the clusters identified above by RSEC. We used the R package {\tt slingshot} \cite{street2018slingshot}. Figure~\ref{fig:lin} illustrates the inferred lineages for the OE dataset, in a two-dimensional space obtained by applying multi-dimensional scaling (MDS) algorithm to the factor scores learned by hGNB. There are three branches in the inferred lineages, with endpoints located in microvillous (MV), mature olfactory sensory neurons (mOSN), and mature sustentacular (mSUS) cells.

\section*{Conclusions}
We propose a hierarchical Bayesian gamma-negative binomial (hGNB) model for extracting low dimensional representations from single-cell RNA sequencing (scRNA-seq) data. hGNB obviates the need for explicit modeling of the zero-inflation prevalent in scRNA-seq count data. Our hGNB can naturally account for covariate effects at both the gene and cell levels, and does not require the commonly adopted pre-processing steps such as normalization. By taking advantage of sophisticated data augmentation techniques, hGNB possesses efficient closed-form Gibbs sampling update equations.
Our experimental results on real-world scRNA-seq data demonstrates that hGNB is capable of identifying insightful cell clusters, especially in complex settings.


\begin{backmatter}

\section*{Competing interests}
  The authors declare that they have no competing interests.

\section*{Author's contributions}
    Conceived the method: SZD, MZ, XQ. Developed the algorithm: SZD, MZ, XQ. Performed the simulations: SZD. Analyzed the results and wrote the paper: SZD, PdF, SHS, MZ, XQ. All authors read and approved the final manuscript.
    

\section*{Acknowledgements}
  This work was supported by the National Science Foundation (NSF) Grants 1553281 and 1812641. The authors also would like to thank Texas A\&M High Performance Research Computing 
  and Texas Advanced Computing Center for providing computational resources to perform experiments in this paper.

\bibliographystyle{bmc-mathphys} 
\bibliography{bmc_article}      

\end{backmatter}
\end{document}